\documentclass[aps, prb, twocolumn, titlepage, longbibliography,showpacs,superscriptaddress]{revtex4-2}
\usepackage{amsmath,amssymb,amsfonts,bm}
\usepackage{graphicx}
\usepackage{epstopdf}
\usepackage{dcolumn}
\usepackage{mathrsfs}
\usepackage{bbold}
\usepackage{dsfont}
\usepackage{float}
\usepackage[colorlinks=true,linkcolor=blue,citecolor=blue, urlcolor=blue,bookmarks=false]{hyperref}
\usepackage{url}
\usepackage{makecell}
\usepackage{tgtermes}
\usepackage{multirow}
\usepackage{natbib}
\usepackage{accents}
\usepackage[position=t,singlelinecheck=off]{subfig}
\usepackage{caption}
\usepackage{ulem}
\usepackage{hyphenat}

\begin{document}

\title{High Curie temperature in diluted magnetic semiconductors (B, Mn)X (X $=$ N, P, As, Sb)}
\date{\today }
\author{Xiang Li}
\affiliation{Kavli Institute for Theoretical Sciences, University of Chinese Academy of Sciences, Beijing 100049, China}
\author{Jia-Wen Li}
\affiliation{Kavli Institute for Theoretical Sciences, University of Chinese Academy of Sciences, Beijing 100049, China}
\author{Jing-Yang You}
\affiliation{Peng Huanwu Collaborative Center for Research and Education, Beihang University, Beijing 100191, China}
\author{Gang Su}\email{gsu@ucas.ac.cn}
\affiliation{Kavli Institute for Theoretical Sciences, University of Chinese Academy of Sciences, Beijing 100049, China}
\affiliation{Institute of Theoretical Physics, Chinese Academy of Sciences, Beijing, China}
\affiliation{Physical Science Laboratory, Huairou National Comprehensive Science Center, Beijing 101400, China}
\affiliation{School of Physical Sciences, University of Chinese Academy of Sciences, Beijing 100049, China}
\author{Bo Gu}\email{gubo@ucas.ac.cn}
\affiliation{Kavli Institute for Theoretical Sciences, University of Chinese Academy of Sciences, Beijing 100049, China}
\affiliation{Physical Science Laboratory, Huairou National Comprehensive Science Center, Beijing 101400, China}
\begin{abstract}
	Doping nonmagnetic semiconductors with magnetic impurities is a feasible way to obtain 
	diluted magnetic semiconductors (DMSs).
	It is generally accepted that for the most extensively studied DMS, (Ga, Mn)As, its highest Curie temperature 
	T$_{\text{C}}$ was achieved at 200 K with a Mn concentration of approximately 16\% in experiments. 
	A recent experiment reported record-breaking high electron 
	and hole mobilities in the semiconductor BAs [\href{https://www.science.org/doi/10.1126/science.abn4290}{Science 377, 437 (2022)}].   
	Since BAs shares the same zinc-blende structure with GaAs, here we predict four DMSs (B, Mn)X (X $=$ N, P, As, Sb) 
	by density functional theory calculations.  
	Using a rescaling method to diminish the overestimation of Curie temperature, our results
	indicate that a significantly higher T$_{\text{C}}$ in the range of 467 K to 485 K for (B, Mn)As with a Mn concentration of around 15.6\%
	and even higher T$_{\text{C}}$ values above the room temperature for (B, Mn)P with a Mn concentration exceeding 9.4\%. 
	Furthermore, using the method of Ab initio Scattering and Transport (AMSET) with first-principles material parameters, we have predicted a 
	hole mobility of 48.9 cm$^{\text{2}}$V$^{\text{-1}}$s$^{\text{-1}}$ at 
	300 K for (B, Mn)As with the hole concentration of about n$_{\text{h}}$ $=$ 4.2 $\times$ 10$^{\text{19}}$ cm$^{\text{-3}}$, 
	which is about two times larger than the hole mobility 
	at 300 K in the calculations for (Ga, Mn)As.
	The hole mobility of (B, Mn)As can be enhanced faster than that of
	(Ga, Mn)As when the hole concentration is decreased. 
	Our findings predict the emergence of a new family of DMS, (B, Mn)X, 
	and are expected to stimulate both experimental 
	and theoretical studies of the DMS with possible high T$_{\text{C}}$.   
\end{abstract}

\maketitle

\section{Introduction}
Magnetic semiconductors,
owing to their combination of 
electron and spin degrees of freedom, 
hold significant promise for spintronic applications.
One effective method to introduce long-range magnetic order and realize
diluted magnetic semiconductor (DMS) is doping magnetic impurities 
such as Cr, Mn, Fe and Co in nonmagnetic semiconductors.
Several efforts have focused on doping  
III-V zinc-blende semiconductors.
In the case of the classic DMS (Ga, Mn)As, a Curie temperature T$_{\text{C}}$ of 110 K 
was obtained with a Mn impurity concentration of 
5.3\% \cite{GaMnAs-DMS-Tc110K-Mn5.3-Ohno}, 
and a higher T$_{\text{C}}$ of 200 K was attained with a Mn impurity concentration of 16\% 
by using non-eqilibrium techniques \cite{GaMnAs-DMS-Tc200K-record}.
However, further enhancing T$_{\text{C}}$ of (Ga, Mn)As becomes challenging 
due to the valance mismatch of Mn$^{\text{2+}}$ and 
Ga$^{\text{3+}}$. This leads Mn impurities to occupy interstitial positions 
as the doping concentration increases \cite{Review-DMS-III-V-semiconductors,Review-DMS-III-V-semiconductors-2,
Review-Ohno-Spintronics,Review-DMS-DFT,DMS-Review-Ohno}.
In experiments of some Mn-doped III-V semiconductors, 
T$_{\text{C}}$ was reported at 60 K for (Ga, Mn)P with a Mn concentration 
of 6\% \cite{GaMnP-DMS-LowTc-60K} 
and T$_{\text{C}}$ reached 15 K in (Ga, Mn)Sb with 
a Mn concentration of 3.9\% \cite{GaMnSb-DMS-LowTc15K}, 
as shown in TABLE \ref{tab: EXP-GaX(X=P,As,Sb)}. 
It is noted that the carrier in DMSs (Ga, Mn)X (X $=$ P, As, Sb) is hole, i.e. p-type. 
The ferromagnetism in (Ga, Mn)N remains a subject of debate,
some experiments reported the room-temperature ferromagnetism 
\cite{GaMnN-DMS-RoomTc-1,GaMnN-DMS-RoomTc-2,GaMnN-DMS-RoomTc-3}
and attributed the high T$_{\text{C}}$ to Mn$_{\text{x}}$N$_{\text{y}}$
clusters instead of dopants \cite{GaMnN-DMS-RoomTc-1,GaMnN-DMS-RoomTc-2},
while other experiments observed low T$_{\text{C}}$ below 10 K 
\cite{GaMnN-DMS-LowTc-1,GaMnN-DMS-LowTc-2,GaMnN-DMS-LowTc-3}.
\begin{table}[H]
	\centering
  \caption{Experimental results of Curie temperature, impurity concentration 
  and carrier types for Mn-doped magnetic
	semiconductors  (Ga, Mn)X (X $=$ P, As, Sb) with zinc-blende structure.}\label{tab: EXP-GaX(X=P,As,Sb)}
	\begin{tabular}{cccc ccc}
		\hline\hline\noalign{\smallskip}
		\makecell[c]{Diluted magnetic \\ semiconductors (years)} 
    & \makecell[c]{Curie \\ temperature T$_{\text{C}}$}
    & \makecell[c]{Impurity \\ concentration} 
		& \makecell[c]{Carrier \\ types} \\
		\noalign{\smallskip}\hline\noalign{\smallskip}
    (Ga, Mn)P (2005) \cite{GaMnP-DMS-LowTc-60K}  & 60 K& 6\% & p-type \\
    (Ga, Mn)As (2011) \cite{GaMnAs-DMS-Tc200K-record} & 200 K& 16\% & p-type \\
    (Ga, Mn)Sb (2014) \cite{GaMnSb-DMS-LowTc15K} & 15 K& 3.9\% & p-type \\
    \noalign{\smallskip}\hline\hline
	\end{tabular}
\end{table}
 
Some DMSs with high T$_{\text{C}}$ have been reported
in experiments in the past decade \cite{EXP-n-GaN-SpinInjection}.
A T$_{\text{C}}$ of 230 K was obtained in p-type 
DMS (Ba, K)(Zn, Mn)$_{\text{2}}$As$_{\text{2}}$ with a Mn 
impurity concentration of 15\% \cite{BaKZn2Mn2As2-FM-DMS-Tc230K-1,BaZn2Mn2As2-FM-DMS-Tc180K}.
This DMS has the advantage 
of decoupled charge and spin doping 
\cite{DMS-EXP-CaKZnMn2As2-DecoupledSpinandElectronDoping,
DMS-EXP-Ba1-xNaxFZn1-xMnxSb-JOS-DecoupledSpinandElectronDoping}.
The hole-mediated ferromagnetism in (Ba, K)(Zn, Mn)$_{\text{2}}$As$_{\text{2}}$
has been discussed in photoemission spectroscopy 
experiments \cite{BaKZn2Mn2As2-FM-DMS-Tc230K-2,BaKZn2Mn2As2-FM-DMS-Tc230K-3} and 
theoretical calculations \cite{BaKZnMnAs-FM-DMS-Theory,BaKZnMnAs-FM-DMS-DFT}, 
similar to the picture discussed in (Ga, Mn)As. 
The T$_{\text{C}}$ of 45 K in Co-doped n-type
DMS Ba(Zn, Co)$_{\text{2}}$As$_{\text{2}}$ was also reported in the experiment \cite{BaZnCoAs-DMS-n-type-Tc45K}.
Experiments have reported  T$_{\text{C}}$ of 340 K in (Ga, Fe)Sb with a
Fe concentration of 25\% \cite{GaFeSb-DMS-Tc340K-Fe25-EXP} and 
T$_{\text{C}}$ of 385 K in (In, Fe)Sb with Fe 
concentration of 35\% \cite{InFeSb-DMS-Tc385K-Fe35-EXP}. 
The valence match between dopant Fe$^{\text{3+}}$ and hosts Ga$^{\text{3+}}$ and In$^{\text{3+}}$
and the very high impurity concentration appear to be
key factors contributing to these high T$_{\text{C}}$ values \cite{DFT+QMC-DMS-Cr-doped-InSb-InAs-GaSb}.    
T$_{\text{C}}$ of Mn-doped Si$_{\text{0.25}}$Ge$_{\text{0.75}}$ 
with a Mn concentration of 5\% was 
reported to be 280 K \cite{Mn-SiGe-thinfilm-HighTc} and 
honeycomb structure (Zn, Co)O monolayer with a Co concentration of 11.1\% was confirmed 
to exhibit long-range ferromagnetism with T$_{\text{C}}$ above
300 K \cite{ZnCoO-DMS-Tc300K-Co11.1-EXP}.

Intrinsic two-dimensional (2D) magnetic semiconductors 
have also been synthesized sucessfully and studied in recent years. 
However, T$_{\text{C}}$ of these 2D magnetic semiconductors in experiments, such as 
CrI$_\text{3}$ \cite{CrI3-FMsemiconductor-Tc45K}, 
Cr$_\text{2}$Ge$_\text{2}$Te$_\text{6}$ \cite{Cr2Ge2Te6-FMsemiconductor-Tc30K},
CrCl$_\text{3}$ \cite{CrCl3-FMsemiconductor-Tc17K}, 
CrBr$_\text{3}$ \cite{CrBr3-FMsemiconductor-Tc34K},
Cr$_\text{2}$S$_\text{3}$ \cite{Cr2S3-FMsemiconductor-Tc75K-1,Cr2S3-FMsemiconductor-Tc75K-2},
CrSBr \cite{CrSBr-FMsemiconductor-Tc146K}
and CrSiTe$_\text{3}$ \cite{CrSiTe3-FMsemiconductor-Tc80K}
are far below the room temperature. 
In order to enhance T$_{\text{C}}$ of 2D magnetic semiconductors, 
effects of strain, electric field and heterostructure 
are theoretically studied, where some high T$_{\text{C}}$ 
2D magnetic semiconductors are proposed 
\cite{CrI3-Substrate-TcEnhancement,
PtBr3-FM-QAHE,
Cr2Ge2Te6-FMsemiconductor-EnhancementTc-Strain,
Cr2Ge2Te2+PtSe2-FM-TcEnhancement,
MnBi2Te4-FM-ElectricField-TcEnhancement,
2Dmagnet-Tc-Review,
MagneticSemiconductor-Review-GroupPaper,
Cr3O6-FM-Semiconductor-DFT,
Review-EXP-StrainedDSM-JOS,
DFT+Model-P-orbit-MagneticTI-QAHE,
Ir2TeI2-monolayer-Strain-FM-MAE}.

For the magnetic semiconductor, high mobility is also 
required for semiconductor applications. 
For the classic DMS (Ga, Mn)As with hole carriers, 
the experimental hole mobility of (Ga, Mn)As is about 10
cm$^{\text{2}}$V$^{\text{-1}}$s$^{\text{-1}}$ 
at room temperature \cite{GaMnAs-mobility-low-1, GaMnAs-mobility-low-2, 
EXP-GaMnAs-Mnconcentration-Holeconcentration-Mobility, EXP-GaMnAs-GAs-mobilities-Fig}.
In fact, there are few magnetic semiconductors with both high T$_{\text{C}}$ and 
high hole mobility in experiments \cite{Mn-SiGe-thinfilm-HighTc}. 
Recently, it was discovered that the semiconductor boron arsenide (BAs), 
isostructural to the zinc-blende GaAs,
has both high electron mobility and hole mobility 1600 
cm$^{\text{2}}$V$^{\text{-1}}$s$^{\text{-1}}$ \cite{
BAs-mobility-1600}, which is much higher than the hole mobility 450 
cm$^{\text{2}}$V$^{\text{-1}}$s$^{\text{-1}}$ of GaAs \cite{GaAs-Hole-Mobility}.
Although BAs had been synthesized
several decades ago \cite{BAs-Wide-Gap-Semiconductor-1,
BAs-Wide-Gap-Semiconductor-2,BAs-Wide-Gap-Semiconductor-3,BAs-Wide-Gap-Semiconductor-4},
some theoretical works have reported the wide gap, 
high electron and hole mobilities and 
high thermal conductivity \cite{BAs-DFT-1,BAs-DFT-2,BAs-DFT-3,
BAs-DFT-4,BAs-DFT-5,BAs-DFT-mobility-1,BAs-DFT-thermal-1,BAs-DFT-thermal-2},
there have been several experiments reporting 
transports of the high electron mobility \cite{BAs-thermal-1,BAs-thermal-2,BAs-thermal-3}, 
the high hole mobility has not been confirmed until 2022 \cite{BAs-mobility-1600}.  
Inspired by the experimental high hole mobility in BAs
and the similarity between semiconductors BAs and GaAs, 
is it possible to obtain the DMS of Mn-doped BAs with both high T$_{\text{C}}$ and high mobility?

In this paper, we carry out calculations on the Mn-doped zinc-blende 
BX (X $=$ N, P, As, Sb) with the density functional theory (DFT).
Our results show a high T$_{\text{C}}$ in the range of 467 K to 485 K in (B, Mn)As with a Mn concentration of 15.6\%, 
higher than the T$_{\text{C}}$ of 200 K in (Ga, Mn)As with a Mn concentration of 16\% in experiment, 
which might be attributed to the shorter length of Mn-As bond and thus stronger Mn-As hybridization in BAs. Mn impurities in BAs keep it 
a p-type semiconductor, and a hole mobility of 48.9 cm$^{\text{2}}$V$^{\text{-1}}$s$^{\text{-1}}$ at 300 K is found in (B, Mn)As 
with a hole concentration of about n$_{\text{h}}$ $=$ 4.2 $\times$ 10$^{\text{19}}$ cm$^{\text{-3}}$.
Other DMSs (B, Mn)N, (B, Mn)P and (B, Mn)Sb are also explored,
T$_{\text{C}}$ values above the room temperature are 
predicted for (B, Mn)P 
with a Mn concentration of above 9.4\% 
while low T$_{\text{C}}$ is obtained in (B, Mn)Sb. 
We use two methods including mean-field
theory and Monte Carlo method for calculations 
of Curie temperatures, results obtained from two methods all support our 
findings on high T$_{\text{C}}$ in DMS of (B, Mn)X family.

\section{Computational details}
First-principles calculations in this work were performed
with the projector augmented wave (PAW) method \cite{PAWmethod}
based on the DFT as implemented
in the Vienna ab initio simulation package (VASP)
\cite{VASP}. The choice of the electron exchange-correlation functional was 
generalized gradient approximation (GGA)
with the form of Perdew-Burke-Ernzerhof (PBE) realization \cite{PBE}.
Lattice constants and atomic positions were fully
relaxed until the maximum force acting on all atoms was less than 1 $\times$ 10$^{\text{-4}}$ eV 
(1 $\times$ 10$^{\text{-3}}$ for exchange coupling calculations)
and the total energy was converged to 1 $\times$ 10$^{\text{-8}}$ eV 
(1 $\times$ 10$^{\text{-7}}$ for exchange coupling calculations)
with the Gaussian smearing method. Calculations of exchange coupling $J_{\text{i}}$ were performed by using
the 3 $\times$ 3 $\times$ 3 conventional cubic supercell. 
The Monkhorst-Pack k-point mesh \cite{M-PKpoints} of size 3 $\times$ 3 $\times$ 3 was used for
the Brillouin zone (BZ) sampling in structure optimization and self-consistent processes of 
exchange coupling calculations, while
for calculations of band structures and mobilities using the primitive cell, 
the Monkhorst-Pack k-point mesh of size 9 $\times$ 9 $\times$ 9 was used.
The plane-wave cutoff energy was set to be 500 eV. 
The electron correlation of the 3d transition atom Mn was considered by using the DFT+U method 
introduced by Dudarev et al. \cite{LDAUTYPE=2}.
All results were obtained with U $=$ 5 eV.
The Monte Carlo simulations were performed with Heisenberg model.
The 10 $\times$ 10 $\times$ 10 supercells with periodic boundary conditions and 
magnetic sites ranging from 2000 to 5000 were adopted
to simulate different doping concentrations. Each temperature calculation used 10$^{\text{4}}$ Monte Carlo
steps to achieve equilibrium \cite{MonteCarloMethod1, MonteCarloMethod2}.

\section{Results}
\subsection{Electronic structure, formation energy and cluster formation of (B, Mn)As}
\begin{figure}[H]
	\captionsetup[figure]{name={FIG.},labelformat=simple,labelsep=period,singlelinecheck=off}
	\centering
	(a)\subfloat{\includegraphics[width=0.5\columnwidth]{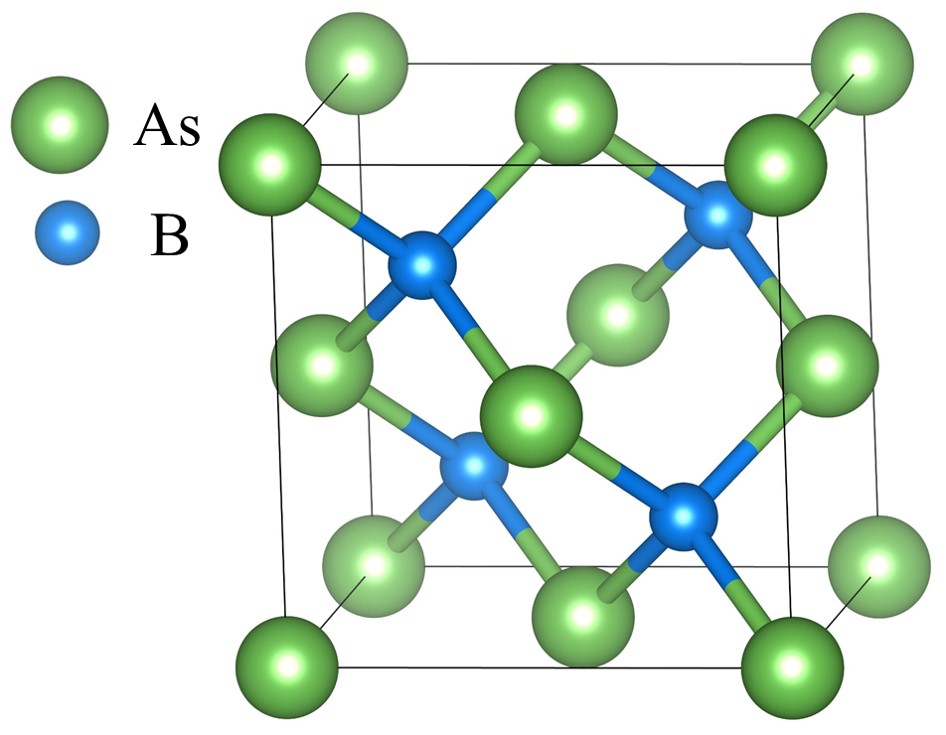}\label{fig: Lattice-BAs}}
	(b)\subfloat{\includegraphics[width=0.4\columnwidth]{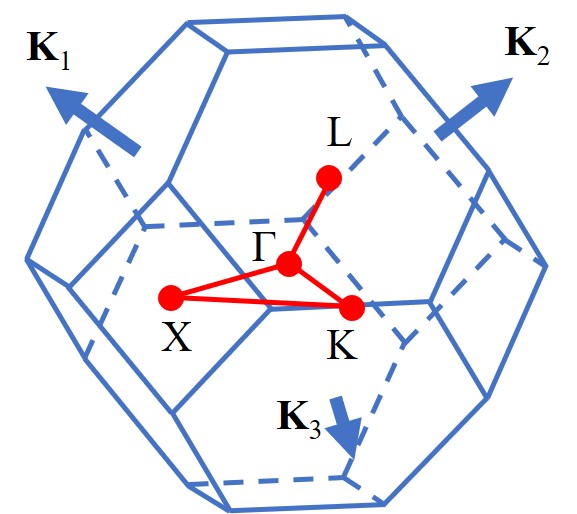}\label{fig: FirstBZ-BAs}}\\
	(c)\subfloat{\includegraphics[width=0.8\columnwidth]{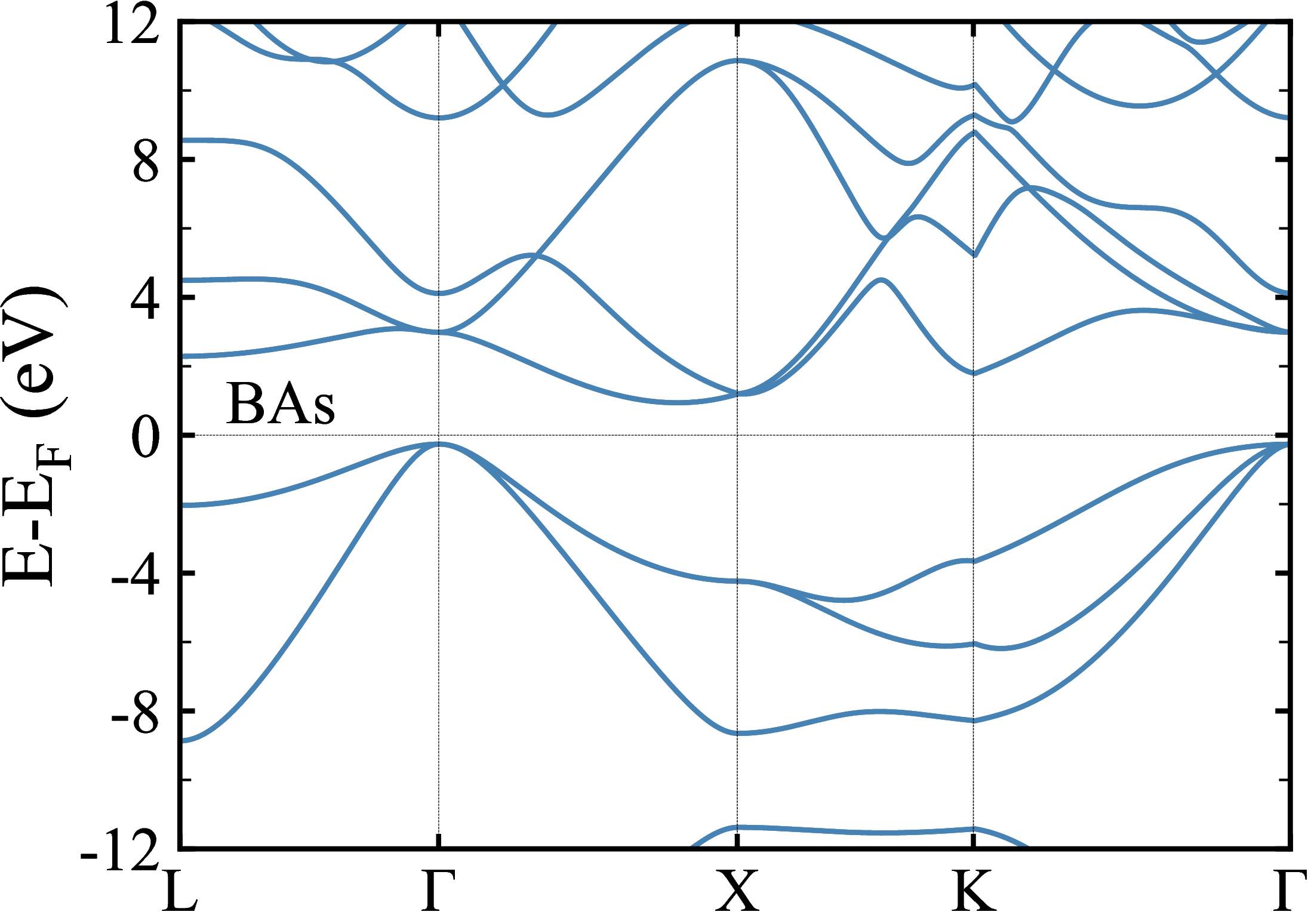}\label{fig: Band-BAs}}\\
	(d)\subfloat{\includegraphics[width=0.8\columnwidth]{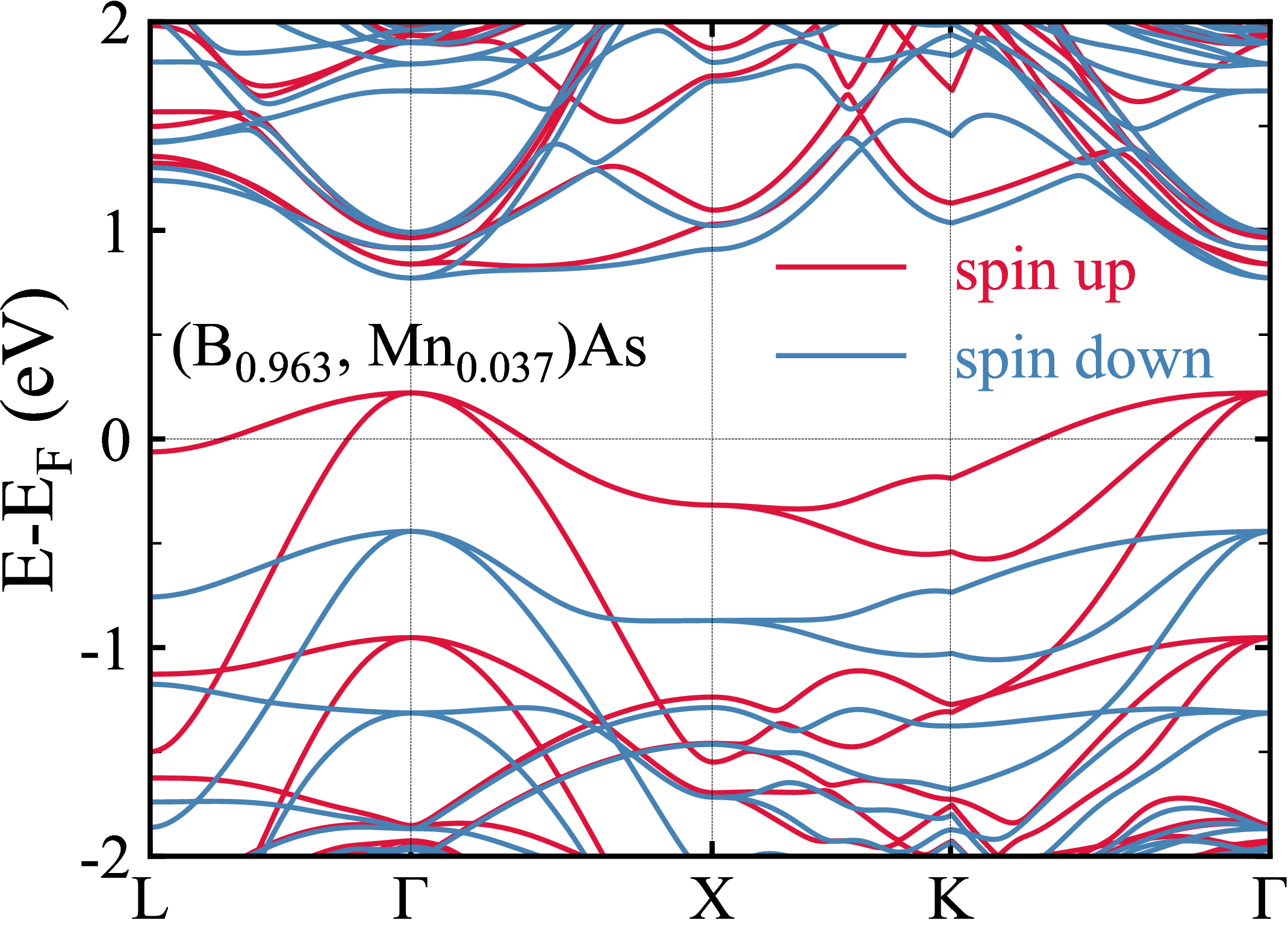}\label{fig: Band-BMnAs}}
	\caption{(a) Zinc-blende crystal structure of the semiconductor BAs and (b) its first Brillouin
	zone. (c) Band structure of BAs with an indirect gap of 1.20 eV. 
	(d) Band structure of the Mn-doped diluted magnetic semiconductor 
	(B$_{\text{0.963}}$, Mn$_{\text{0.037}}$)As with a direct gap of 0.55 eV at $\Gamma$.}
\end{figure}

Zinc-blende BAs has the space group F$\bar{\text{4}}$3m (No. 216),
its crystal structure and the first Brillouin zone with high symmetry paths indicated with red color 
are shown in FIG. \ref{fig: Lattice-BAs} and FIG. \ref{fig: FirstBZ-BAs}, respectively. 
As shown in FIG. \ref{fig: Band-BAs}, BAs 
is a semiconductor with a calculated indirect gap of 1.20 eV, which is consistent with 
experimental values of 1.46 eV \cite{BAs-Wide-Gap-Semiconductor-1,BAs-Wide-Gap-Semiconductor-2,BAs-Wide-Gap-Semiconductor-3,BAs-Wide-Gap-Semiconductor-4} 
and other calculations \cite{BAs-DFT-1,BAs-DFT-2,BAs-DFT-3,BAs-DFT-4,BAs-DFT-5}.  
The 3 $\times$ 3 $\times$ 3 primitive supercell with B$_{\text{26}}$MnAs$_{\text{27}}$ 
is adopted to study the electronic structure of the Mn-doped BAs.
As shown in FIG. \ref{fig: Band-BMnAs}, (B$_{\text{0.963}}$, Mn$_{\text{0.037}}$)As is a p-type semiconductor and
the band gap shrinks to 0.55 eV and becomes direct at $\Gamma$.

The formation energy of (B, Mn)As is higher than that of (Ga, Mn)As  
for As-rich and Mn-rich condition and both of them 
are positive which suggests low solubility of Mn impurities. See Supplemental Material for more details \cite{Supplement}. 
However, non-equilibrium 
growth technologies like metal-organic chemical vapor deposition (MOCVD) or molecular beam epitaxy (MBE) etc. 
can effectively raise the doping concentration in the diluted magnetic semiconductors to a high level, 
which is much away from the solid solubility limit \cite{EXP-ImpurityConcentration-AwayFromSolubility}. 
For (Ga, Mn)As, Mn as high as 20\% can be successfully incorporated into GaAs by optimizing low-temperature MBE growth condition 
without inducing phase separation \cite{EXP-GaMnAs-High-dopingconcentration-MBE,
EXP-GaMnAs-High-dopingconcentration-MBE-2,EXP-GaMnAs-High-dopingconcentration-MBE-3, EXP-GaMnAs-High-dopingconcentration-MBE-4}.
For other examples, large defect formation energies (ranging from 3.7 eV to 6.95 eV \cite{DFT-ZnO-interstitialZn-formationenergy}) 
of interstitial Zn atoms in ZnO are reported,
however, ZnO with high concentration of interstitial Zn atoms are prepared based on non-equilibrium growth \cite{EXP-ZnO-high-interstitialZn}.
Non-equilibrium growth indicates that kinetic
barriers may preserve the high-energy defects once them form even for impurity 
concentration exceeding the nominal equilibrium estimated values. 
Considering non-equilibrium growth above, it is thus expected that 
Mn concentration in (B, Mn)As can be also reached to a 
high level as in (Ga, Mn)As.
It is noted that the interstitial Mn and Mn 
aggregation have important roles in (Ga, Mn)As, experimental studies demonstrate 
a reduction in the Curie temperature T$_{\text{C}}$ due to clusters of Mn 
dopants in (Ga, Mn)As \cite{EXP-GaMnAs-cluster, EXP-GaMnAs-cluster-2, EXP-GaMnAs-cluster-3}.
For (Ga, Mn)As, the tetramer cluster is found to be 
energetically most favorable. The calculations suggest that Mn atoms have a strong 
tendency to form tetramer cluster in (Ga, Mn)As, which is consistent with experiments and other calculations
\cite{Supercell-DFT-DMS-5, GaMnAs-FormationEnergy,  
DFT-GaMnAs-cluster-tendency, DFT-GaMnAs-cluster-tendency-2, DFT-GaMnAs-cluster-tendency-3, DFT-GaMnAs-cluster-tendency-4, DFT-GaMnAs-cluster-tendency-5}.
For (B, Mn)As, the Mn configuration with two dimers is energetically most favorable. 
The results suggest that tendency of cluster for Mn in (B, Mn)As should be smaller than that in (Ga, Mn)As.
Details can be found in Supplemental Material \cite{Supplement}.
For simplicity and purpose of obtaining DMSs with higher Curie temperature T$_{\text{C}}$, 
here we only study substitutional Mn impurities in (B, Mn)As.

\subsection{Exchange couplings of (B, Mn)As}
\begin{figure}[H]
	\captionsetup[subfigure]{labelformat=simple}
	\centering
	\subfloat{\includegraphics[width=1\columnwidth]{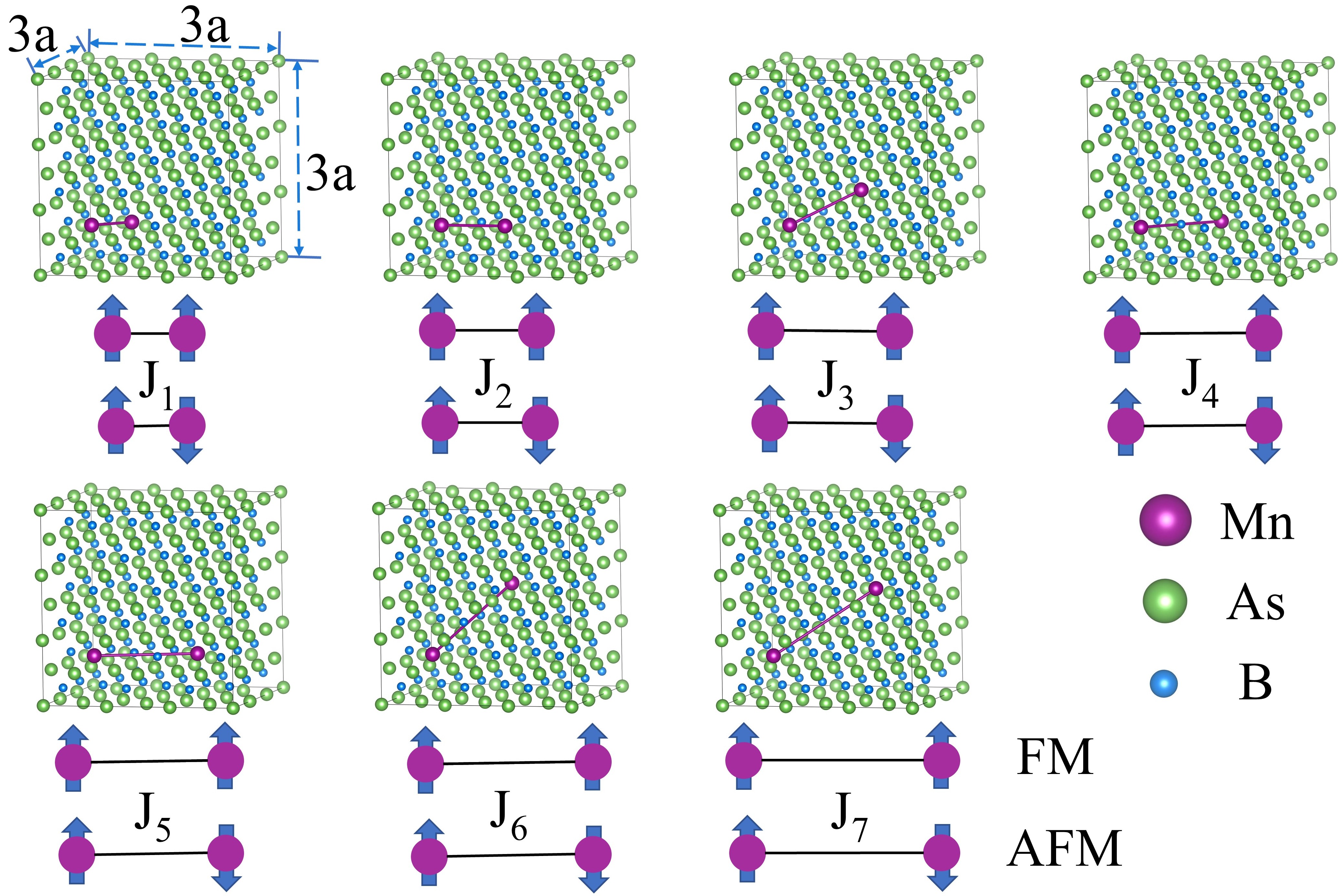}}
	\caption{Constructed supercells to calculate 
	the exchange coupling parameters between Mn impurities in semiconductor BAs.  
	Seven different supercells of B$_{\text{106}}$Mn$_{\text{2}}$As$_{\text{108}}$ are used, where the
	exchange couplings $J_{\text{i}}$ with $\text{i}=1$-$7$ denoting the i-th nearest neighbors
	are calculated independently. See text for details.}\label{fig: Jmethod}
\end{figure}

In order to estimate T$_{\text{C}}$ of the Mn-doped BAs, we should calculate
the magnetic exchange coupling of two Mn impurities first. 
Long-range magnetic interactions between far distant Mn 
impurity sites are considered to establish magnetic order.
The disordered magnetic system is mapped to the classical Heisenberg-type Hamiltonian with 
$J_{\text{i}}$ ($\text{i}=1$-$7$) denoting the i-th
nearest-neighbor exchange coupling
between two impurity sites, respectively, 
Hamiltonian is expressed as
\begin{align}
	H =&-J_{\text{1}} \sum_{i<j} \bm{S}_{i} \cdot \bm{S}_{j}
	-J_{\text{2}} \sum_{i<j} \bm{S}_{i} \cdot \bm{S}_{j}
	-J_{\text{3}} \sum_{i<j} \bm{S}_{i} \cdot \bm{S}_{j}\cr
	&-J_{\text{4}} \sum_{i<j} \bm{S}_{i} \cdot \bm{S}_{j}
	-J_{\text{5}} \sum_{i<j} \bm{S}_{i} \cdot \bm{S}_{j}
	-J_{\text{6}} \sum_{i<j} \bm{S}_{i} \cdot \bm{S}_{j}\cr
	&-J_{\text{7}} \sum_{i<j} \bm{S}_{i} \cdot \bm{S}_{j}
\end{align}
where two impurities are coupled ferromagnetically for conditions of $J_{\text{i}}$$>$0 while 
antiferromagnetically for conditions of $J_{\text{i}}$$<$0.
In this work, exchange coupling parameters
are calculated via the method illustrated in FIG. \ref{fig: Jmethod}.
To avoid the influence of unphysical
exchange interaction from magnetic atoms in mirror neighbor cells and reduce the computation,
some different 3 $\times$ 3 $\times$ 3 cubic supercells are
adopted as shown in FIG. \ref{fig: Jmethod}, where a stands for the lattice constant 
of BAs.

Seven different supercells of B$_{\text{106}}$Mn$_{\text{2}}$As$_{\text{108}}$ are used.  
Ferromagnetic (FM) configuration with parellel spins and antiferromagnetic (AFM) configuration 
with anti-parellel spins structures are considered.
This method is the conventional way to construct supercells in past studies 
of DMSs \cite{ExchangeCoupling-DFT-FM-AFM-1,ExchangeCoupling-DFT-FM-AFM-2,
Supercell-DFT-DMS-1,Supercell-DFT-DMS-2,Supercell-DFT-DMS-3,Supercell-DFT-DMS-4,Supercell-DFT-DMS-5,DMS-3dDopedMoS2-DFT,
DMS-MnDopedMoS2-TMDs,DMS-FM-TMDs-DFT,DMS-CoDopedphosphorene-DFT,DMS-FeDopedSnS2-DFT,DMS-FeDopedSnS2-DFT+EXP,
DASP-DopedSystem-Software}.
$J_{\text{1}}$, $J_{\text{2}}$, $J_{\text{3}}$, $J_{\text{4}}$, $J_{\text{5}}$, $J_{\text{6}}$ and $J_{\text{7}}$ 
are calculated independently.
For $J_{\text{1}}$, two impurities are located at (1/3, 1/6, 5/6)3a and (1/2, 1/6, 2/3)3a
illustrated in FIG. \ref{fig: Jmethod} and 
their energy expressions are given by
\begin{align}\label{eq: case-J1}
	E_{\text{AFM}}=&E_{0}+J_{1}S^{2},\cr
	E_{\text{FM}}=&E_{0}-J_{1}S^{2},
\end{align}    
where $E_{0}$ is the energy independent of spin configurations.
The energy difference between FM and AFM configurations gives the $J_{\text{1}}$.
For $J_{\text{2}}$, two impurities are located at (1/3, 1/6, 5/6)3a and (2/3, 1/6, 5/6)3a and
their energy expressions are given by
\begin{align}\label{eq: case-J2}
	E_{\text{AFM}}=&E_{0}+J_{2}S^{2},\cr
	E_{\text{FM}}=&E_{0}-J_{2}S^{2}.
\end{align}
$J_{\text{2}}$ can be calculated by Eq. \eqref{eq: case-J2} with DFT results of $E_{\text{AFM}}$ and $E_{\text{FM}}$. 
For $J_{\text{3}}$, two impurities are located at (1/3, 1/6, 5/6)3a and (2/3, 1/3, 2/3)3a and 
their energy expressions are given by
\begin{align}\label{eq: case-J3}
	E_{\text{AFM}}=&E_{0}+J_{3}S^{2},\cr
	E_{\text{FM}}=&E_{0}-J_{3}S^{2}.
\end{align}
$J_{\text{3}}$ can be obtained by Eq. \eqref{eq: case-J3}. 
For $J_{\text{4}}$, two impurities are located at (1/3, 1/6, 5/6)3a and (2/3, 1/6, 1/2)3a and 
their energy expressions are given by
\begin{align}\label{eq: case-J4}
	E_{\text{AFM}}=&E_{0}+J_{4}S^{2},\cr
	E_{\text{FM}}=&E_{0}-J_{4}S^{2}.
\end{align}
$J_{\text{4}}$ can be obtained by Eq. \eqref{eq: case-J4}. 
For $J_{\text{5}}$, two impurities are located at (1/3, 1/6, 5/6)3a and (5/6, 1/3, 5/6)3a and 
their energy expressions are given by
\begin{align}\label{eq: case-J5}
	E_{\text{AFM}}=&E_{0}+J_{5}S^{2},\cr
	E_{\text{FM}}=&E_{0}-J_{5}S^{2}.
\end{align}
$J_{\text{5}}$ can be obtained by Eq. \eqref{eq: case-J5}. 
For $J_{\text{6}}$, two impurities are located at (1/3, 1/6, 5/6)3a and (2/3, 1/2, 1/2)3a and 
their energy expressions are given by
\begin{align}\label{eq: case-J6}
	E_{\text{AFM}}=&E_{0}+J_{6}S^{2},\cr
	E_{\text{FM}}=&E_{0}-J_{6}S^{2}.
\end{align}
$J_{\text{6}}$ can be obtained by Eq. \eqref{eq: case-J6}. 
For $J_{\text{7}}$, two impurities are located at (1/3, 1/6, 5/6)3a and (5/6, 1/2, 2/3)3a and 
their energy expressions are given by
\begin{align}\label{eq: case-J7}
	E_{\text{AFM}}=&E_{0}+2J_{7}S^{2},\cr
	E_{\text{FM}}=&E_{0}-2J_{7}S^{2}.
\end{align}
$J_{\text{7}}$ can be obtained by Eq. \eqref{eq: case-J7}. 

The obtained exchange coupling parameters 
$J_{\text{i}}$
for Mn-doped BAs and GaAs 
as a function of distance between two impurities R$_{\text{i}}$ (in
unit of the lattice constant a)
are shown in TABLE \ref{tab: J-GaAs-BAs-U=5eV} and FIG. \ref{fig: J-GaAs-BAs-U5}, where U $=$ 5 eV.
For Mn atom with five local d electrons, the spin is 
set to be S $=$ 5/2. 
It is noted that obtained values of exchange couplings $J_{\text{i}}$ are independent of 
impurity concentrations, and the disorder effect 
due to impurities in DMS is included in coordination numbers and configuration average
in the mean-field approximation of T$_{\text{C}}$ in Eq. \eqref{eq: Meanfield} 
and Monte Carlo method with Eq. \eqref{eq: MCaverage}. 
All magnetic couplings are ferromagnetic for both (B, Mn)As and (Ga, Mn)As.
The nearest coupling $J_{\text{1}}$ dominants the exchange interaction. 
$J_{\text{i}}$ drops drastically when two impurities are placed more distant,
the trend of $J_{\text{i}}$ as a function of the distance of two impurity sites 
shows Ruderman-Kittel-Kasuya-Yosida (RKKY) type of behavior \cite{Review-DMS-DFT}. 
Long-range magnetic couplings can connect distant magnetic atoms in diluted case,
which is responsible for ferromagnetic order and Curie temperature.
$J_{\text{1}}$ of (B, Mn)As is much larger than that of (Ga, Mn)As,
which might be attributed to shorter Mn-As bond length 
(d$_{\text{Mn-As}}$ $=$ 2.552 $\mathring{\text{A}}$ for (Ga, Mn)As and 2.358 $\mathring{\text{A}}$ for (B, Mn)As) and 
stronger Mn-As hybridization in (B, Mn)As
\cite{Review-DMS-III-V-semiconductors, Review-DMS-DFT, DFT-GaMnAs-FM-mechanism, 
EXP-GaMnAs-FM-pdinteracyopm-mechanism, EXP-GaMnAs-pdexchange}.

\begin{table}[H]
	\centering
	\caption{Exchange coupling parameters $J_{\text{i}}$ (meV) of two Mn impurities in (B, Mn)As and (Ga, Mn)As
    calculated by DFT+U with spin S $=$ 5/2 and Hubbard correlation parameter U $=$ 5 eV.
	$J_{\text{1}}$-$J_{\text{7}}$ represent the 1st through the 7th nearest-neighbor 
	exchange couplings, respectively.}\label{tab: J-GaAs-BAs-U=5eV}
	\begin{tabular}{cccc ccccc}
		\hline\hline\noalign{\smallskip}
		& & \makecell[c]{$J_{\text{1}}$ \\ (meV)} & \makecell[c]{$J_{\text{2}}$ \\ (meV)} 
		& \makecell[c]{$J_{\text{3}}$ \\ (meV)} & \makecell[c]{$J_{\text{4}}$ \\ (meV)} 
		& \makecell[c]{$J_{\text{5}}$ \\ (meV)} & \makecell[c]{$J_{\text{6}}$ \\ (meV)} 
		& \makecell[c]{$J_{\text{7}}$ \\ (meV)}\\
		\noalign{\smallskip}\hline\noalign{\smallskip}
		&\makecell[c]{(B, Mn)As} & \makecell[c]{19.036}& \makecell[c]{4.312}& \makecell[c]{3.754}
		& \makecell[c]{10.135}& \makecell[c]{0.801}& \makecell[c]{4.565}& \makecell[c]{2.117}\\
		\noalign{\smallskip}\hline\noalign{\smallskip}
		&\makecell[c]{(Ga, Mn)As} & \makecell[c]{9.525}& \makecell[c]{1.545}& \makecell[c]{1.975}
		& \makecell[c]{2.854}& \makecell[c]{0.269}& \makecell[c]{1.931}& \makecell[c]{0.877}\\
		\noalign{\smallskip}\hline\hline
	\end{tabular}
\end{table}

\subsection{Curie temperatures of (B, Mn)As}\label{sec: MFATc}
Curie temperature T$_{\text{C}}$ is calculated by the
mean-field approximation (MFA) in Eq. \eqref{eq: Meanfield} and by the Monte Carlo (MC) method in Eq. \eqref{eq: MCaverage}, respectively.
For the MFA method,
\begin{align}\label{eq: Meanfield}
	T^{\text{MFA}}_{\text{C}}
	=&\frac{2}{3k_{\text{B}}}S(S+1)\sum_{\text{i}}\sum_{\alpha}Z_{\text{i}}^{\alpha}P_{\alpha}J_{\text{i}}, 
\end{align}
$Z_{\text{i}}^{\alpha}$ is the coordination number of one doping configuration $\alpha$ for the given doping concentration,
$P_{\alpha}$ is the probability of this doping configuration.
Doping configurations $\alpha$ are generated via Disorder code \cite{Disorder-1,Disorder-2}.
To simulate the doped system with random distribution of magnetic impurities, 
different doping configurations for a given
doping concentration are considered in
the 2 $\times$ 2 $\times$ 2 cubic supercell with B$_{\text{32}}$As$_{\text{32}}$.
The condition of the doping concentration 6.25\% is given as an example where 
two B atoms in the supercell B$_{\text{32}}$As$_{\text{32}}$ are replaced with Mn impurities, so
there are two inequivalent doping configurations and their corresponding $Z_{\text{i}}^{\alpha}$  
and $P_{\alpha}$ are listed in TABLE \ref{tab: 6.25-P-Z-alpha}.
All mean-field values are calculated with the fixed spin S $=$ 5/2.
All doping configurations of considered doping concentrations 
host exchange couplings between all magnetic sites 
even for diluted case, which forms the three-dimensional (3D) magnetic network. 
3D periodic magnetic sublattice is constructed and corresponds to  
each doping configuration with the specific probability,
see Supplemental Material 
for more details about this method and coordination numbers of other doping concentrations \cite{Supplement}.
For MC method in Eq. \eqref{eq: MCaverage}, T$_{\text{C}}^{\alpha}$ 
for each doping configuration is calculated by MC simulation on corresponding magnetic sublattice $\alpha$
with exchange couplings listed in TABLE \ref{tab: J-GaAs-BAs-U=5eV}, T$^{\text{MC}}_{\text{C}}$ is the 
configuration average of T$_{\text{C}}^{\alpha}$ for a fixed doping concentration.
\begin{align}\label{eq: MCaverage}
	T^{\text{MC}}_{\text{C}}
	=&\sum_{\alpha}P_{\alpha}T_{\text{C}}^{\alpha}.
\end{align}

\begin{figure}[H]
	\captionsetup[figure]{name={FIG.},labelformat=simple,labelsep=period,singlelinecheck=off,format=plain}
	\centering
	(a)\hspace{20pt}\subfloat{\includegraphics[width=0.85\columnwidth]{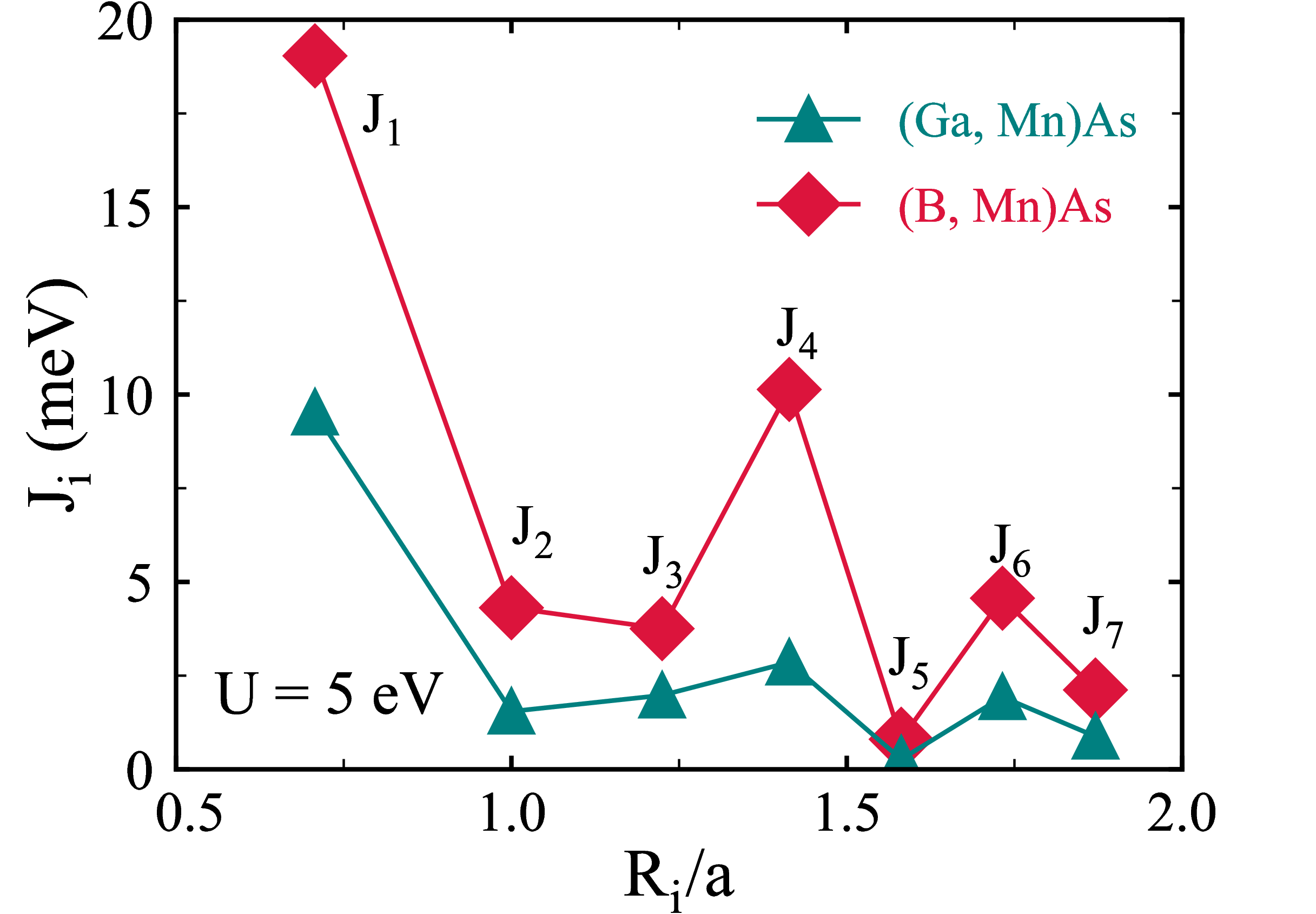}\label{fig: J-GaAs-BAs-U5}}\\
	(b)\hspace{20pt}\subfloat{\includegraphics[width=0.85\columnwidth]{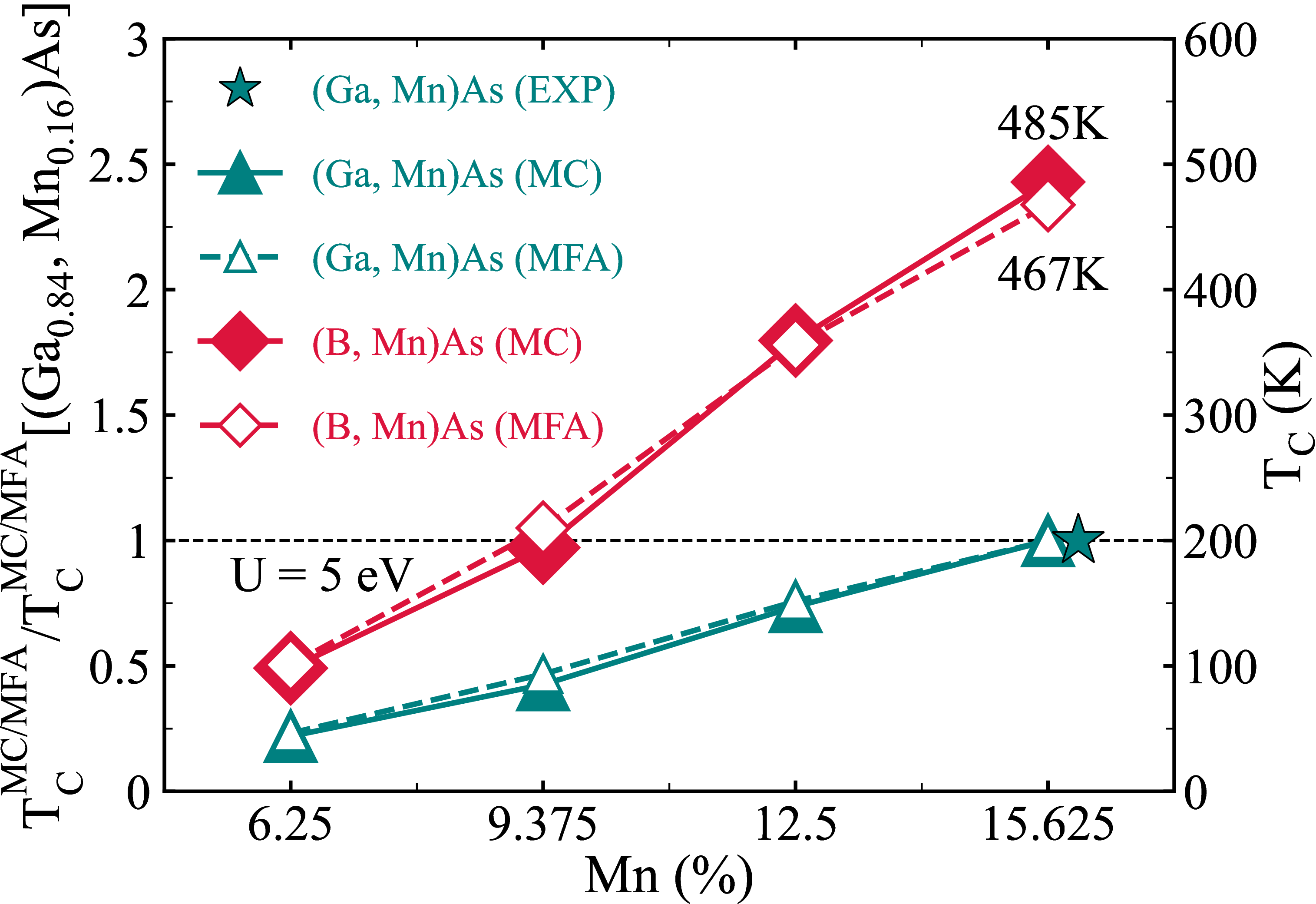}\label{fig: Tc-GaAs-BAs-U5}}\\
	\caption{Exchange coupling parameters $J_{\text{i}}$ and Curie temperature T$_\text{C}$ for Mn-doped BAs and GaAs. 
	(a) Exchange coupling $J_{\text{i}}$ as a function of distance 
	R$_{\text{i}}$ between two Mn impurities, where a is the lattice constant and Hubbard 
	correlation parameter U $=$ 5 eV. The same data in TABLE \ref{tab: J-GaAs-BAs-U=5eV}.
	(b) The ratio of Curie temperatures T$_{\text{C}}^{\text{MFA}}$/T$_{\text{C}}^{\text{MFA}}$[(Ga$_{\text{0.84}}$, Mn$_{\text{0.16}})$As] 
	calculated by
	the mean-field approximation (MFA) in Eq. \eqref{eq: Meanfield}
	and T$_{\text{C}}^{\text{MC}}$/T$_{\text{C}}^{\text{MC}}$[(Ga$_{\text{0.84}}$, Mn$_{\text{0.16}})$As] 
	by the Monte Carlo (MC) simulation in Eq. \eqref{eq: MCaverage}
	as a function of Mn impurity concentration, where U $=$ 5 eV. 
	Both T$_{\text{C}}^{\text{MFA}}$[(Ga$_{\text{0.84}}$, Mn$_{\text{0.16}})$As] 
	and T$_{\text{C}}^{\text{MC}}$[(Ga$_{\text{0.84}}$, Mn$_{\text{0.16}})$As]
	correspond to the T$^{\text{EXP}}_\text{C}$ $=$ 200 K of (Ga$_{\text{0.84}}$, Mn$_{\text{0.16}})$As 
	in experiment, respectively, which is depicted by a black dashed line.
	The estimated T$_\text{C} = $ 485 K by MC method and T$_\text{C} = $ 467 K by 
	MFA method for (B, Mn)As with Mn concentration 15.625\% are labeled.}\label{fig: J-GaAs-BAs}
\end{figure}
For Mn concentration 15.625\%, T$_{\text{C}}^{\text{MFA}}$ and T$_{\text{C}}^{\text{MC}}$ are calculated with 
up to the 7-th nearest-neighbor exchange couplings listed in TABLE \ref{tab: J-GaAs-BAs-U=5eV} and by the 
average of 215 doping configurations.
The calculated T$_{\text{C}}^{\text{MFA}}$[(Ga$_{\text{0.84}}$, Mn$_{\text{0.16}})$As] $=$ 2355 K 
and T$_{\text{C}}^{\text{MC}}$[(Ga$_{\text{0.84}}$, Mn$_{\text{0.16}})$As] $=$ 635 K
are far beyond the experimental value T$_{\text{C}}^{\text{EXP}}$[(Ga$_{\text{0.84}}$, Mn$_{\text{0.16}})$As] $=$ 200 K.
In order to provide an improved estimation of T$_{\text{C}}$ for
(Ga, Mn)As and (B, Mn)As, we analyze the calculated T$_{\text{C}}^{\text{MFA}}$ 
and T$_{\text{C}}^{\text{MC}}$ in the following way. 
As shown in FIG. \ref{fig: Tc-GaAs-BAs-U5}, 
the ratio of Curie temperatures T$_{\text{C}}^{\text{MFA}}$/T$_{\text{C}}^{\text{MFA}}$[(Ga$_{\text{0.84}}$, Mn$_{\text{0.16}})$As] 
by the MFA method in Eq. \eqref{eq: Meanfield} and 
T$_{\text{C}}^{\text{MC}}$/T$_{\text{C}}^{\text{MC}}$[(Ga$_{\text{0.84}}$, Mn$_{\text{0.16}})$As] 
by the MC simulation in Eq. \eqref{eq: MCaverage}
as a function of Mn impurity concentration is plotted, where U $=$ 5 eV. 
Both T$_{\text{C}}^{\text{MFA}}$[(Ga$_{\text{0.84}}$, Mn$_{\text{0.16}})$As]
and T$_{\text{C}}^{\text{MC}}$[(Ga$_{\text{0.84}}$, Mn$_{\text{0.16}})$As]
are taken as the unit one for MFA and MC methods, respectively, 
which physically corresponds to the experimental value
T$_{\text{C}}^{\text{EXP}}$[(Ga$_{\text{0.84}}$, Mn$_{\text{0.16}})$As] $=$ 200 K \cite{GaMnAs-DMS-Tc200K-record}. 
We plot such correspondence in FIG. \ref{fig: Tc-GaAs-BAs-U5} by a black dashed line. 
Both MC and MFA methods show 
similar trends of Curie temperatures of (Ga, Mn)As and (B, Mn)As after rescaling as shown in Fig. \ref{fig: Tc-GaAs-BAs-U5}.
The high Curie temperature T$_{\text{C}}$ $=$ 485 K by MC method 
and T$_{\text{C}}$ $=$ 467 K by MFA method are predicted for (B, Mn)As with Mn concentration 15.625\% for (B, Mn)As,
larger than experimental 
value of Curie temperature T$_{\text{C}}^{\text{EXP}}$ of 200 K for (Ga, Mn)As 
with a Mn concentration of 16\% \cite{GaMnAs-DMS-Tc200K-record}.
\begin{table}[H]
	\centering
	\caption{For doping concentration 6.25\%, doping configuration $\alpha$, coordination number $Z_{\text{i}}^{\alpha}$
	of the exchange coupling $J_{\text{i}}$, and probability $P_{\alpha}$. }\label{tab: 6.25-P-Z-alpha}
	\begin{tabular}{cccc ccc}
		\hline\hline\noalign{\smallskip}
		$\alpha$ &$Z_{\text{i}}^{\alpha}$ for Mn1 &$Z_{\text{i}}^{\alpha}$ for Mn2 & $P_{\alpha}$  \\
		\noalign{\smallskip}\hline\noalign{\smallskip}
		1 & \{0, 0, 0, 0, 0, 8, 0\} & \{0, 0, 0, 0, 0, 8, 0\} & 1/13 \\ 
		2 & \{0, 0, 2, 0, 0, 0, 4\} & \{0, 0, 2, 0, 0, 0, 4\} & 12/13 \\ 
		\noalign{\smallskip}\hline\hline
	\end{tabular}
\end{table}

\subsection{Curie temperatures of (B, Mn)X (X $=$ N, P, Sb)}\label{sec: BMnXandGaMnX}
In order to verify the reliability of the above rescaling method,
by the MC simulation, the ratio of Curie temperature 
T$_{\text{C}}^{\text{MC}}$/T$_{\text{C}}^{\text{MC}}$[(Ga$_{\text{0.84}}$, Mn$_{\text{0.16}})$As] 
as a function of the Mn impurity concentration for 
(Ga, Mn)P and (Ga, Mn)Sb are calculated
and shown in FIG. \ref{fig: Tc-GaX-U5} with exchange couplings listed in Supplemental Material \cite{Supplement}. Experimental 
Curie temperatures of (Ga, Mn)P, (Ga, Mn)As and (Ga, Mn)Sb from TABLE \ref{tab: EXP-GaX(X=P,As,Sb)} are labeled.
By the rescaling method, T$_{\text{C}}^{\text{EXP}}$ $=$ 65 K is 
obtained for (Ga, Mn)P with a Mn concentration of 6\%, 
which is consistent with the experiment \cite{GaMnP-DMS-LowTc-60K}, 
as shown in FIG. \ref{fig: Tc-GaX-U5}. 
Ultra low T$_{\text{C}}^{\text{EXP}}$ is obtained for 
(Ga, Mn)Sb with a very low Mn concentration by our rescaling method, 
which is also consistent with the experiment \cite{GaMnSb-DMS-LowTc15K}, 
as shown in FIG. \ref{fig: Tc-GaX-U5}. 
Thus, it is expected that T$_{\text{C}}^{\text{EXP}}$ obtained 
in FIG. \ref{fig: Tc-GaX-U5} and FIG. \ref{fig: Tc-BX-U5}
by the rescaling method could diminish the overestimation of T$_{\text{C}}^{\text{MFA}}$ 
and T$_{\text{C}}^{\text{MC}}$
and give a reasonable estimation of T$_{\text{C}}$. 
Curie temperatures of possible DMSs (B, Mn)X with X $=$ N, P, Sb are also calculated in the same way. 
Band structures of (B, Mn)N, (B, Mn)P and (B, Mn)Sb suggest that they are all p-type semiconductors
as shown in Supplemental Material \cite{Supplement}.
FIG. \ref{fig: Tc-BX-U5} predicts high T$_{\text{C}}$ values above room temperature 
for (B, Mn)P when the Mn concentration is above 9.375\%
and low T$_{\text{C}}$ for (B, Mn)Sb.  
Similar results are obtained for mean-field method with U $=$ 5 eV as shown in Supplemental Material \cite{Supplement}. 
\begin{figure}[H]
	\captionsetup[subfigure]{labelformat=simple}
	\centering
	(a)\subfloat{\includegraphics[width=0.9\columnwidth]{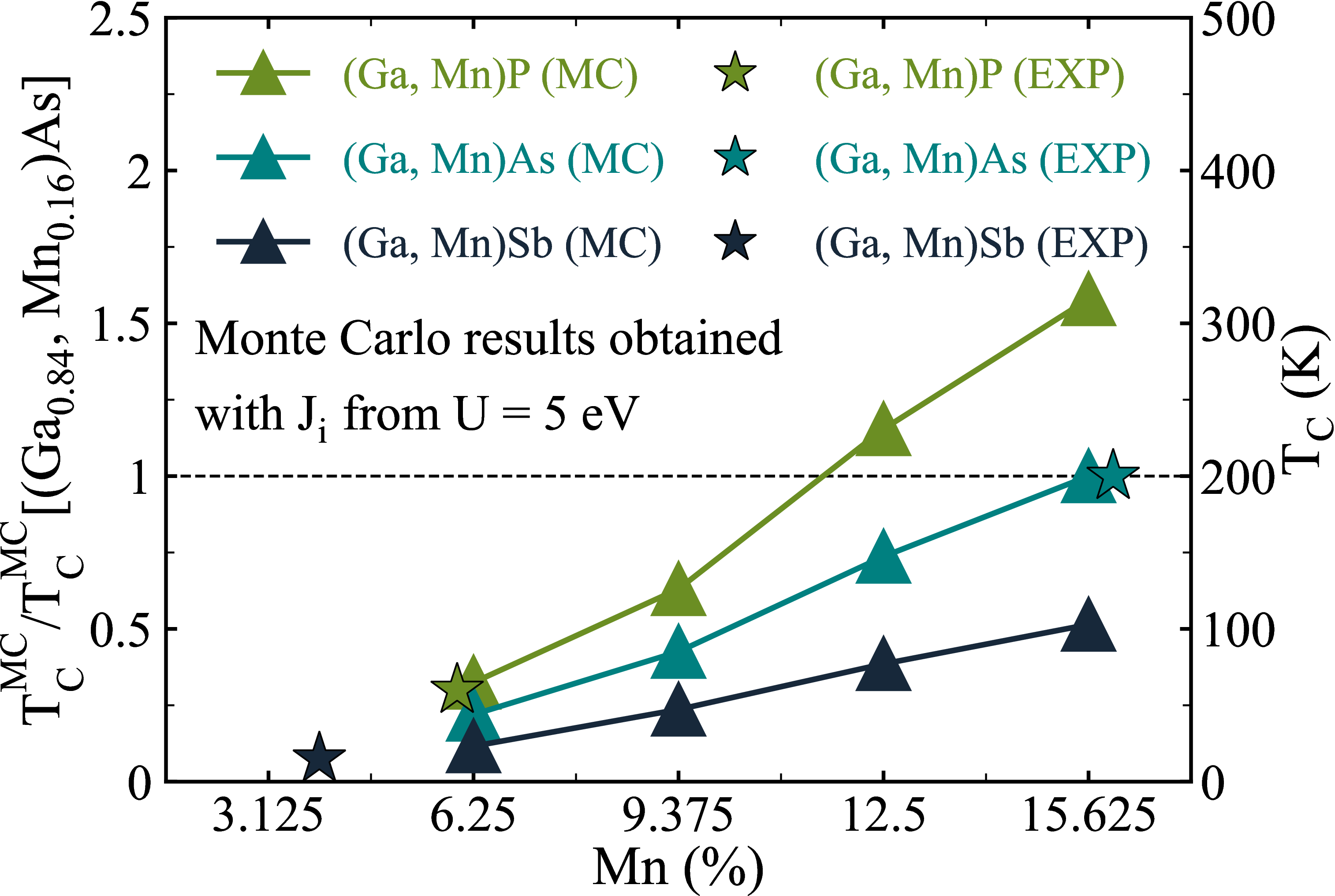}\label{fig: Tc-GaX-U5}}\\
	(b)\subfloat{\includegraphics[width=0.9\columnwidth]{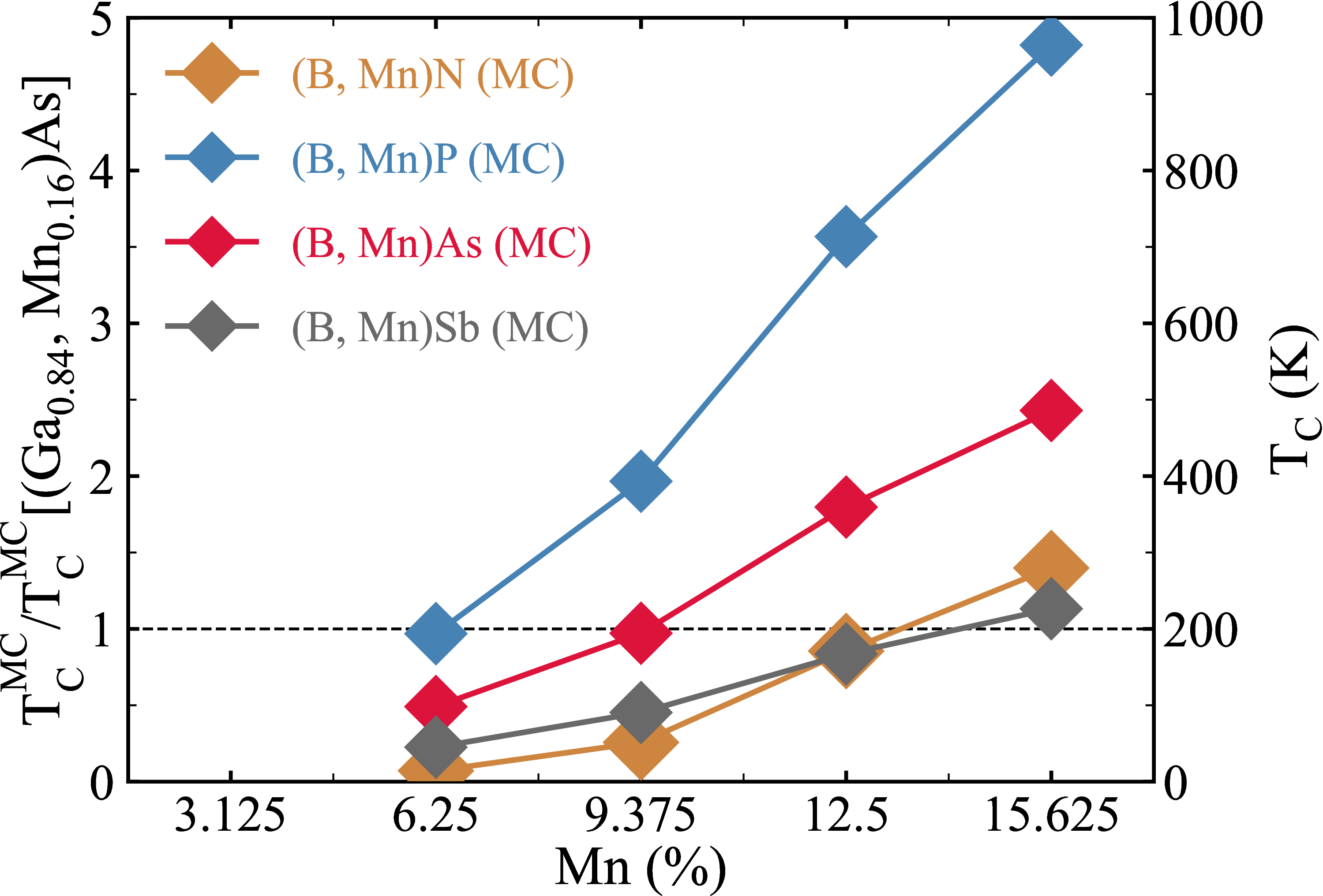}\label{fig: Tc-BX-U5}}\\
	\caption{By the Monte Carlo (MC) method in Eq. \eqref{eq: MCaverage}, the ratio of Curie temperature 
	T$_{\text{C}}^{\text{MC}}$/T$_{\text{C}}^{\text{MC}}$[(Ga$_{\text{0.84}}$, Mn$_{\text{0.16}})$As] 
	as a function of Mn impurity concentration for 
	(a) Mn-doped GaX (X $=$ P, As Sb), and (b) Mn-doped BX (X $=$ N, P, As, Sb). Magnetic exchange couplings with U $=$ 5 eV is considered.  
	T$_{\text{C}}^{\text{MC}}$[(Ga$_{\text{0.84}}$, Mn$_{\text{0.16}})$As] corresponds to 
	T$_{\text{C}}^{\text{EXP}}$ of 200 K of (Ga$_{\text{0.84}}$, Mn$_{\text{0.16}})$As in experiment \cite{GaMnAs-DMS-Tc200K-record}. 
	Such correspondence is depicted by a black dashed line in (a) and (b).}\label{fig: Tc-GaX-BX-U5}
\end{figure}
\begin{figure*}[htbp]
	\captionsetup[subfigure]{labelformat=simple}
	\centering
	(a)\hspace{5pt}\subfloat{\includegraphics[width=0.9\columnwidth]{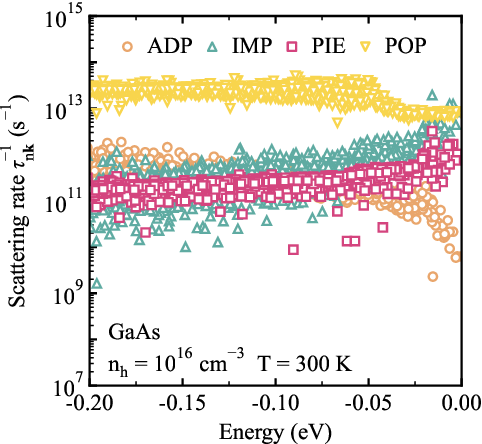}\label{fig: Rates-GaAs}}
	(b)\hspace{5pt}\subfloat{\includegraphics[width=0.9\columnwidth]{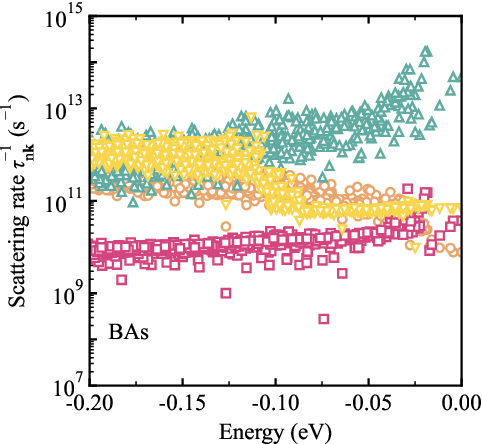}\label{fig: Rates-BAs}}\\
	(c)\hspace{5pt}\subfloat{\includegraphics[width=0.9\columnwidth]{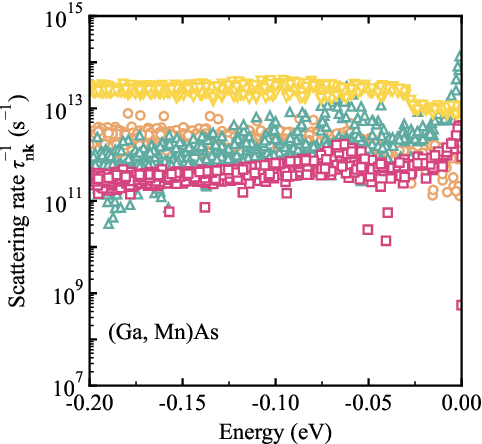}\label{fig: Rates-GaMnAs}}
	(d)\hspace{5pt}\subfloat{\includegraphics[width=0.9\columnwidth]{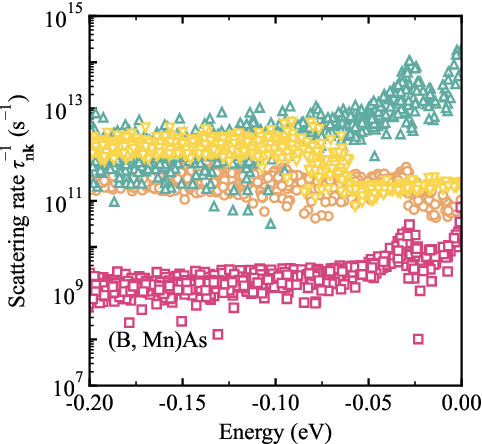}\label{fig: Rates-BMnAs}}\\
	\captionsetup{justification=raggedright, singlelinecheck=false}
	\caption{Calculated scattering rates $\tau_{n\bm{k}}^{\text{-1}}$ (s$^{\text{-1}}$) of (a) GaAs, (b) BAs, (c) (Ga, Mn)As and (d) (B, Mn)As at 300 K 
	with hole concentration n$_{\text{h}}$ $=$ 10$^{\text{16}}$ cm$^{\text{-3}}$. 
	Zero energy corresponds to the valence band maximum. 
	Scatter points at one energy level are from different energy bands $n$ and $\bm{k}$, which are summed for calculations of hole mobilities in FIG. \ref{fig: HoleMobility}.
	Four scattering mechanisms in semiconductors including acoustic deformation potential scattering (ADP), 
	ionized impurity scattering (IMP), piezoelectric scattering (PIE) and polar optical phonon scattering (POP) are considered while calculating scattering
	rates and hole mobilities. \arraybackslash}\label{fig: Rates}
\end{figure*}

\subsection{Hole mobility of (B, Mn)As}
\begin{table*}[htbp]
	\centering
	\captionsetup{justification=raggedright, singlelinecheck=false}
	\caption{Experimental and calculated lattice properties of GaAs, BAs, (Ga, Mn)As and (B, Mn)As. C$_{\text{11}}$,
	C$_{\text{12}}$ and C$_{\text{44}}$ are elastic constants (GPa), 
	$\omega_{\text{PO}}$ is polar-optical phonon frequency (THz),
	$\varepsilon_{\infty}$ and $\varepsilon_{\text{0}}$ are high-frequency dielectric constant (eV) and static dielectric constant (eV) respectively
	and e$_{\text{14}}$ is piezoelectric constant (C/m$^{\text{2}}$).}\label{tab: properties}
	\begin{tabular}{ccccc cccc}
		\hline\hline\noalign{\smallskip}
		& \makecell[c]{GaAs \\(EXP)}
		& \makecell[c]{GaAs\\ (DFT)}
		& \makecell[c]{GaAs \\(this work)}
		& \makecell[c]{BAs \\(EXP)}
		& \makecell[c]{BAs\\ (DFT)}  
		& \makecell[c]{BAs \\(this work)}  
		& \makecell[c]{(Ga, Mn)As \\ (this work)}
		& \makecell[c]{(B, Mn)As \\(this work)}\\
		\noalign{\smallskip}\hline\noalign{\smallskip}
		C$_{\text{11}}$ & 118.41 \cite{GaAs-Elastic-EXP} & 106.50 \cite{GaAs-Elastic-DFT}& 114.95 &285.00 \cite{BAs-properties-EXP-Elastic}& 275.80 \cite{BAs-properties-DFT}& 266.11 &86.77 & 230.74\\
		C$_{\text{12}}$ & 53.70 \cite{GaAs-Elastic-EXP} & 60.20	\cite{GaAs-Elastic-DFT}& 42.42 &79.50	\cite{BAs-properties-EXP-Elastic} & 73.30 \cite{BAs-properties-DFT}& 64.68 & 47.60& 71.98\\
		C$_{\text{44}}$ & 59.12 \cite{GaAs-Elastic-EXP} & 33.60	\cite{GaAs-Elastic-DFT}& 57.21 &149.00 \cite{BAs-properties-EXP-Elastic}&	168.70	\cite{BAs-properties-DFT}& 144.34 & 30.32& 114.10\\
		$\omega_{\text{PO}}$ & 8.76 \cite{GaAs-PolarOptical-EXP} &8.69 \cite{GaAs-PolarOptical-DFT}&	7.52 & 21.42 \cite{BAs-PolarOptical-EXP}	& 20.80 \cite{BAs-properties-DFT}& 20.37 & 6.25& 14.57\\
		$\varepsilon_{\infty}$& 10.89 \cite{GaAs-EXP-StaticDie-book} &	13.35 \cite{GaAs-DFT-StaticDie-14}	& 10.46	&	& 9.91	\cite{BAs-properties-DFT}&  9.84 & & 11.09\\
		$\varepsilon_{\text{0}}$& 12.90 \cite{GaAs-EXP-StaticDie-book} & 14.94 \cite{GaAs-DFT-StaticDie-14}	 &	12.63 &	& 10.01	\cite{BAs-properties-DFT}& 9.93 & & 11.26	\\
		e$_{\text{14}}$& -0.16 \cite{GaAs-Piezo-EXP}& -0.12 \cite{GaAs-Piezo-DFT}	&	&  &  & -0.05 & & 0.03	\\
		\noalign{\smallskip}\hline\hline
	\end{tabular}
\end{table*}

In order to estimate the scattering rates
$\tau_{n\bm{k}}^{\text{-1}}$ and hole mobilities $\mu_{\text{h}}$ 
of BAs, GaAs, (B, Mn)As and (Ga, Mn)As at 300 K, we applied the method of Ab initio Scattering and
Transport (AMSET) \cite{AMSET}. For a more thoughtful consideration of transport of carriers in semiconductors,
four important scattering mechanisms are included in the code of AMSET, i.e. 
the acoustic deformation potential scattering (ADP) \cite{DeformationBardeen, Deformation-2, Deformation-3, Deformation-4}, 
ionized impurity scattering (IMP) \cite{Impurity-1, Impurity-2}, piezoelectric scattering (PIE) \cite{Piezoelectric-1, Piezoelectric-2, Piezoelectric-3} 
and polar optical phonon scattering (POP) \cite{PolarPhonon}.
ADP is the scattering between electron and 
longitudinal and transverse acoustic phonon, 
POP is the scattering between electron and polar optical phonon,
PIE is the scattering between electron and long-wavelength acoustic phonon,
IMP is the scattering between electron and ionized impurity.
Four scattering mechanisms make different contributions to the coupling matrix $g_{nm}(\bm{k},\bm{q})$ 
in Eq. \eqref{eq: scatteringrates-elastic} for elastic scatterings including ADP, POP and PIE 
\begin{align}\label{eq: scatteringrates-elastic}
	\tau_{n\bm{k}\to m\bm{k}+\bm{q}}^{-1}=\frac{2\pi}{\hbar}|g_{nm}(\bm{k},\bm{q})|^{2}\delta(\Delta E_{\bm{k},\bm{q}}^{nm})
\end{align}
and Eq. \eqref{eq: scatteringrates-inelastic} for inelastic scattering IMP with Fermi's golden rule
\begin{align}\label{eq: scatteringrates-inelastic}
	\tau_{n\bm{k}\to m\bm{k}+\bm{q}}^{-1}=&\frac{2\pi}{\hbar}|g_{nm}(\bm{k},\bm{q})|^{2}\cr
	&[(n_{\bm{q}}+1-f_{m\bm{k}+\bm{q}}^{0})\delta(\Delta E_{\bm{k,\bm{q}}}^{nm}-\hbar\omega_{\bm{q}})\cr
	&+(n_{\bm{q}}+f_{m\bm{k}+\bm{q}}^{0})\delta(\Delta E_{\bm{k},\bm{q}}^{nm}+\hbar\omega_{\bm{q}})]
\end{align}
where $\tau_{n\bm{k}}$ is the relaxation time for each band and $\bm{k}$ point, 
the $-\hbar\omega_{\bm{q}}$ and $+\hbar\omega_{\bm{q}}$ terms correspond to 
scatterings by emission and absorption of a phonon respectively, 
$n_{\bm{q}}$ is the Bose-Einstein occupation, $f_{n\bm{k}}^{0}$ is the Fermi-Dirac distribution,
$\Delta E_{\bm{k},\bm{q}}^{nm}=E_{n\bm{k}}-E_{m\bm{k}+\bm{q}}$ and $E_{n\bm{k}}$ is the band energy. 
The coupling matrix element is the probability of scattering from an initial state $|n,\bm{k}\rangle$
to final state $|m,\bm{k}+\bm{q}\rangle$ via a phonon with frequency $\omega_{\bm{q}}$ and wave vector $\bm{q}$.
Parameters of GaAs, BAs, (Ga, Mn)As, (B, Mn)As that are related to four scattering mechanisms in AMSET are obtained via first principles calculations and 
are listed in TABLE \ref{tab: properties}. For GaAs, the experimental value of piezoelectric constant is adopted in AMSET.
The dielectric constants of (Ga, Mn)As are determined to be the same 
with the experimental values of GaAs due to not much difference of dielectric functions 
of GaAs and (Ga, Mn)As in experiment \cite{EXP-GaMnAs-dielectrics-exp-1, EXP-GaMnAs-dielectrics-exp-2}, and the piezoelectric tensor 
is also determined to be the same with that of GaAs.
The final scattering rate for each scattering mechanism is obtained by
\begin{align}
	\tau_{n\bm{k}}^{-1}=\sum_{m}\int\frac{\mathrm{d}\bm{q}}{\Omega_{\text{BZ}}}\tau_{n\bm{k}\to m\bm{k}+\bm{q}}^{-1}
\end{align}
and the final mobility can be calculated using the above four scattering rates. 

\begin{figure}[H]
	\captionsetup[subfigure]{labelformat=simple}
	\centering
	(a)\hspace{5pt}\subfloat{\includegraphics[width=0.9\columnwidth]{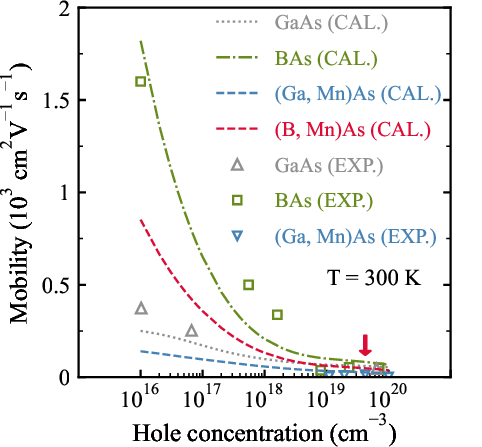}\label{fig: HoleMobility-All}}\\
	(b)\hspace{5pt}\subfloat{\includegraphics[width=0.9\columnwidth]{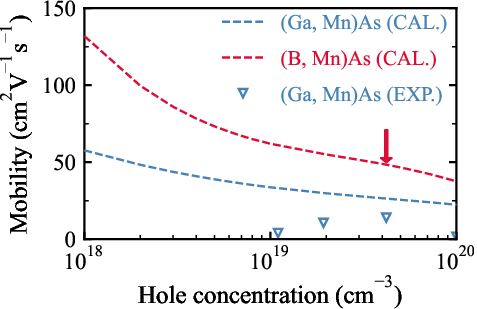}\label{fig: HoleMobility-zoomin}}\\
	\caption{
	The hole mobility as a function of hole concentration for semiconductors GaAs, 
	BAs, (Ga, Mn)As and (B, Mn)As calculated by AMSET code. For comparison, some 
	experimental values of hole mobility for GaAs \cite{GaAs-Hole-Mobility-1, GaAs-Hole-Mobility-2}, BAs \cite{BAs-mobility-1600} and (Ga, Mn)As 
	\cite{GaMnAs-mobility-low-1, GaMnAs-mobility-low-2, 
	EXP-GaMnAs-Mnconcentration-Holeconcentration-Mobility, EXP-GaMnAs-GAs-mobilities-Fig} at 300 K are shown.}\label{fig: HoleMobility}
\end{figure}

Scattering rates for GaAs, BAs and (B, Mn)As 
at 300 K are depicted in FIG. \ref{fig: Rates-GaAs}, \ref{fig: Rates-BAs} and \ref{fig: Rates-BMnAs}
respectively with a fixed hole concentration of n$_{\text{h}}$ $=$ 10$^{\text{16}}$ cm$^{\text{-3}}$, where 
separate scattering rates of ADP, IMP, PIE and POP are labeled by different markers and colors.
Zero energy corresponds to the valence band maximum and
scatter points at one energy level are from different energy bands E$_{n\bm{k}}$, which are summed 
for calculations of hole mobilities in FIG. \ref{fig: HoleMobility-All}.
For polar material GaAs in FIG. \ref{fig: Rates-GaAs}, polar optical phonon scatterings is donimant at 300 K, which is 
consistent with the experimental measurement \cite{GaAs-EXP-Polar-Dominant-1, GaAs-EXP-Polar-Dominant-2}.
Scattering rates from polar optical phonons is near two orders of magnitude larger than contributions from other scattering mechanisms
and thus responsible for its low hole mobility at low hole concentration. The ionized impurity scattering and piezoelectric scattering 
are comparable at the band edge and also contribute to the decay of lifetime of band egde hole.
The acoustic deformation potential scattering decreases when the energy approaches the band edge
and is not dominant in all energy range.

Due to less difference of $\varepsilon_{\infty}$ and $\varepsilon_{\text{0}}$, BAs is weakly ionic \cite{BAs-DFT-5}, 
which leads to evident disparity and relatively weak scatterings of polar optical phonons compared with GaAs as depicted in FIG. \ref{fig: Rates-BAs}.
Moreover, scattering rates of ionized impurities and holes of BAs is dominant and 
is two orders of magnitude larger than GaAs in energy range around the band edge. 
Ionized impurity scattering rates increases and polar optical phonons and acoustic deformation potential scattering rates decreases drastically
when energy level approaches the band edge, which is qualitatively consistent with other calculations \cite{BAs-mobility-1600}.
Ultra-low piezoelectric constant leads to negligible piezoelectric scattering in BAs. 
Calculated total hole mobilities by considering scattering rates from ADP, IMP, PIE and POP depicted in FIG. \ref{fig: HoleMobility-All} are consistent well with 
experimental data of GaAs \cite{GaAs-Hole-Mobility-1, GaAs-Hole-Mobility-2}, BAs \cite{BAs-mobility-1600}
and (Ga, Mn)As \cite{GaMnAs-mobility-low-2} indicated by different markers and colors, which suggests the reliability for mobility calculations in AMSET.
By lowing ionized impurity concentration (hole concentration), hole mobilities of BAs and GaAs can both increase according to FIG. \ref{fig: HoleMobility-All}, 
however, strong polar optical phonons scatterings even at low hole concentration limit 
the value of hole mobility of GaAs while the value of BAs might be enhanced further due to less influence of polar optical phonons \cite{Supplement}.
For scattering rates of (Ga, Mn)As, scatterings between polar optical phonon 
and holes remain high and the ionized impurity scattering and the deformation potential 
scattering are enhanced, which bring down the hole mobility 
compared with GaAs as shown in FIG. \ref{fig: HoleMobility-All}. 
Results in FIG. \ref{fig: Rates-GaMnAs} and FIG. \ref{fig: Rates-BMnAs} suggest the fact of 
suppressed carrier life time and hole mobilities when 
doping magnetic Mn impurities in BAs and GaAs.

For (B, Mn)As in FIG. \ref{fig: Rates-BMnAs}, scatterings from ionized impurities are still dominant. Scatterings from POP and ADP increases 
which can be attributed to the decreasing of frequency of polar optical phonon while scatterings from PIE decrease compared with BAs.
In fact, it is found that ADP is not the dominant scattering mechanism in all energy ranges for holes of GaAs, BAs, (Ga, Mn)As and (B, Mn)As,
which will lead to the overestimation of carrier mobility if only 
ADP is considered. Much weaker phonon scatterings and piezoelectric scattering in (B, Mn)As 
contribute to longer carrier lifetime and thus higher hole mobility than (Ga, Mn)As at the same hole concentration.
Hole mobilities of GaAs, BAs, (Ga, Mn)As and (B, Mn)As are all found to decrease with the increasing of 
hole concentration and the IMP contributes most to the decay of lifetime of carriers at high hole concentration which stresses the nonnegligible impurity effect \cite{Supplement}.
The doping of Mn impurities increases the hole concentration in GaAs and BAs due to valance mismatch of Mn$^{\text{2+}}$ and Ga$^{\text{3+}}$ or B$^{\text{3+}}$, 
which results in the low hole mobility about 10
cm$^{\text{2}}$V$^{\text{-1}}$s$^{\text{-1}}$ at 300 K for the hole concentration 
n$_{\text{h}}$ = 10$^{\text{19}}$ $\sim$ 10$^{\text{20}}$ cm$^{\text{-3}}$ in (Ga, Mn)As  
with its maximum value 14 cm$^{\text{2}}$V$^{\text{-1}}$s$^{\text{-1}}$ for the high hole concentration n$_{\text{h}}$ $=$ 4.2 
$\times$ 10$^{\text{19}}$ cm$^{\text{-3}}$ \cite{GaMnAs-mobility-low-1, GaMnAs-mobility-low-2, EXP-GaMnAs-Mnconcentration-Holeconcentration-Mobility, EXP-GaMnAs-GAs-mobilities-Fig}.
With the high hole concentration n$_{\text{h}}$ $=$ 4.2 
$\times$ 10$^{\text{19}}$ cm$^{\text{-3}}$ 
marked by the red arrow in FIG. \ref{fig: HoleMobility}, the calculated hole 
mobility of (Ga, Mn)As is 26.6 cm$^{\text{2}}$V$^{\text{-1}}$s$^{\text{-1}}$, about two times larger than the experimental value.
With the same hole concentration n$_{\text{h}}$ $=$ 4.2 
$\times$ 10$^{\text{19}}$ cm$^{\text{-3}}$, the calculation predicts a higher hole mobility of 
48.9 cm$^{\text{2}}$V$^{\text{-1}}$s$^{\text{-1}}$ for (B, Mn)As, about two times larger than that of (Ga, Mn)As in calculations. In addition, 
it is predicted that the hole mobility of (B, Mn)As can be enhanced 
faster than that of (Ga, Mn)As when the hole concentration is decreased, as shown in FIG. \ref{fig: HoleMobility}.

\section{Discussion}
Besides a high 
Curie temperature, (B, Mn)As exhibits a direct band gap, 
suggesting favorable optical properties such as 
photoluminescence and magnetic field-controlled 
light intensity. These characteristics hold promise for 
applications in spin field-effect transistors, 
spin light-emitting diodes, 
and non-volatile memory \cite{Review-Spintronics-RMP}.
Some semiconductors such as chalcopyrites including CdGeP$_{\text{2}}$, 
ZnGeP$_{\text{2}}$ and ZnSnAs$_{\text{2}}$ have also been shown to become ferromagnetic
upon Mn doping, with remarkably 
high Curie temperature \cite{DFT-MnDopedchalcopyrites-DMS}, which shows promising feature prospect of DMS.

By forming junctions with superconductors, 
the experimental extraction of spin polarization 
in carriers from (B, Mn)As and interfacial transparency 
becomes achievable. This is possible through 
conductance measurements at voltages below the 
superconductor gap using point-contact Andreev 
reflection spectroscopy 
\cite{Model-NSemicon-SJunction-AndreeReflection-SpinPolarized,
EXP-GaMnAs-SpinPolarization-AndreevSpectroscopy}. 
Furthermore, DMS with spin-active 
interfaces may result in equal-spin reflection, 
contributing to the elusive 
phenomenon of spin-triplet superconductivity
\cite{Review-ProximityEffects-SuperconductorFerromagnet-Junction-RMP}.

For applications of (B, Mn)X in superconductor and valleytronics, 
arrays of magnetic tunneling junctions constructed from the (B, Mn)X family 
on the surface of an s-wave superconductor can generate controllable, 
non-trivial magnetic spin textures that act on the 2D electron 
gas through proximity effects. This process results in effective 
p-wave pairing and the formation of Majorana bound states, paving 
the way for applications in topological 
quantum computing \cite{Model-DMS-Application-MTJs-Superconductor}. 
Furthermore, spin injection from p-type (B, Mn)X 
to n-type transition metal dichalcogenides monolayer might enable 
the higher-temperature control of 
valley polarization \cite{EXP-GaMnAs_TMDs_ValleyPolarization_SpinInjection}. 

For spintronic applications, 
the combination of the p-type (B, Mn)X family 
with electron-doped n-type BX can be employed 
to create a novel p-n junction, p-(B, Mn)X/n-BX, 
exhibiting intriguing spin-voltaic phenomena. 
This configuration holds potential for use in spin 
solar cells and spin photodiodes 
\cite{EXP-DMS-SpinSolarCell-SpinPhotodiodeEffect}. 
Unpolarized light has the ability to induce spin injection 
in the p-region and spin extraction in the n-region, 
eliminating the need for indirect and local methods 
to induce charge current first. This approach helps 
mitigate Joule heating concerns.
Moreover, a bipolar magnetic junction transistor 
with magnetic field-control of the amplification 
will be very interesting 
by organizing p-(B, Mn)X/n-BX/p-BX 
\cite{Model-DMS-BipolarMagneticJunctionTransistor, 
EXP-DMS-BipolarMagneticJunctionTransistor}.

\section{Conclusion}
By using the density functional theory, we have studied 
the Mn-doped zinc-blende semiconductors (B, Mn)X (X $=$ N, P, As, Sb). Our calculations 
show that Mn impurities introduce ferromagnetism in these semiconductors. 
Using a rescaling method to diminish the overestimation of Curie temperature by mean-field approximation,
we predict a high T$_{\text{C}}$ in the range of 467 K to 485 K
for (B, Mn)As with 15.6\% Mn impurities, which is higher than the highest 
T$_{\text{C}}$ of 200 K for (Ga, Mn)As with 16\% Mn impurities in the experiment. 
T$_{\text{C}}$ values above the room temperature are predicted for (B, Mn)P with a Mn concentration of above 9.4\%.
By using AMSET combining semi-empirical theory and first-principles parameters of materials, (B, Mn)As is found to 
have the hole mobility 48.9 cm$^{\text{2}}$V$^{\text{-1}}$s$^{\text{-1}}$
at 300 K which is about two times larger than the hole mobility 26.6 cm$^{\text{2}}$V$^{\text{-1}}$s$^{\text{-1}}$
of (Ga, Mn)As in calculations.
It should be noted that the hole mobility of (B, Mn)As can be enhanced faster than that of
(Ga, Mn)As when the hole concentration is decreased.
Our results highlight the new DMS (B, Mn)As with high T$_{\text{C}}$.

\section{Acknowledgements}
This work is supported by National Key R\&D Program of China (Grant No. 2022YFA1405100), National Natural Science Foundation
of China (Grant No. 12074378), Chinese Academy of Sciences
(Grants No. YSBR-030, No. JZHKYPT-2021-08,
No. XDB33000000).

\bibliographystyle{apsrev4-2}

\end{document}